\documentclass[twocolumn,nofootinbib,amsmath,amssymb,a4paper]{revtex4}

\usepackage{graphicx}
\usepackage{dcolumn}
\usepackage{bm}
\usepackage{multirow}
\usepackage{verbatim}

\newcommand{\ButoKpipi} {\ensuremath{ B^\pm \to K^\pm \pi^+ \pi^- }}
\newcommand{\BdtoKpipi} {\ensuremath{ B^0,\bar{B^0} \to K_S \pi^+ \pi^- }}

\begin{document}

\title{CKM $\gamma$ phase from $B \to K \pi \pi$ }

\author{Ignacio Bediaga, Gabriel Guerrer and Jussara M. de Miranda \\  
	bediaga@cbpf.br, guerrer@cbpf.br and jussara@cbpf.br \\ $\,$ \\}
  
\affiliation{Centro Brasileiro de Pesquisas F\'\i sicas, Rua Xavier Sigaud 150,
 22290-180  -- Rio de Janeiro, RJ, Brazil\\}

\begin{abstract}
We discuss a method to extract the CKM  $\gamma$ angle 
combining Dalitz plot analysis of $B^{\pm} \to K^{\pm} \pi^+ \pi^-$ 
and untagged $B^0$, $\bar B^0 \to K_s \pi^+ \pi^-$. 
The method also allows obtaining the ratio 
and phase difference between the {\it tree} and  {\it penguin} contributions from  
$B^0$ and  $\bar B^0 \to K^{*\pm} \pi^{\mp} $ decays and direct CP asymmetry
between $B^0$ and  $\bar{B^0}$. 
From Monte Carlo studies 
of 100K events for the neutral mesons, we show the possibility of measuring $\gamma$ with a 
precision of $\sim 5^o$.
\end{abstract}

\maketitle

\section{Introduction}
In the  Standard Model, CP violation is implemented by the
existence of CKM complex  parameters. Interference between processes with 
different weak phases  contributing to a same final state,  can generate 
asymmetry on charge conjugate $B$ meson decays, allowing one to measure the weak phase
difference since the asymmetry effects are in some way proportional to it. 
The common method \cite{betababar, betabelle} to extract the CKM $\beta$ phase,  
explores  the  interference between the phases generated in the
$B^0 - \bar B^0$ oscillation, in the decay to the same $J/\Psi K_s$ final state.
For CKM $\gamma$, the most well established methods \cite{ads, glw, ggsz} explores the interference 
in the time-independent decays
$B^- \to \bar{D^{0}} K^{-}$ and $B^- \to D^{0} K^{-}$, 
when $\bar{D^{0}}$ and $D^{0}$ decays to the same final state.
Using these methods,  
$\gamma$ should be determined  at the LHC \cite{ schneider} with precision 
of $\sim 5^0 $, for one year of data taking. 

Beyond two body interferences, one could explore the asymmetry 
associated with the Dalitz interference between intermediary states
in three body $B$ decays.
This was initially proposed for $ B^{\pm} \to \pi^{\pm} \pi^+ \pi^-$ \cite{bbgg},
where $\chi_{c0}$ plays a fundamental role as reference channel.  
However, the method is statistically limited by the low contribution of the
Cabibbo suppressed amplitude  $ B^{\pm} \to \chi_{c0} \pi^{\pm}$, 
estimated on the already observed Cabibbo allowed $B^{\pm}\to \chi_{c0} K^{\pm}$ 
channel \cite{BaBark2pi,  bellek2pi}. 
Recently, new effort is being made to explore
$\gamma$ in $B\to K \pi\pi$ decays.
While some approaches requires time-dependent analysis \cite{cps, gpsz},
we first showed \cite{bgm} how to proceed with an untagged analysis,
with the benefit of higher statistics.
Computer simulations \cite{TDRLHCb} showed that tagging introduces inefficiencies in the order of 90\% in LHCb
data.


\section{Method}
 
The major intermediate sates for \ButoKpipi and \BdtoKpipi 
\cite{BaBark2pi, bellek2pi, bellekspipi}, are resumed in table \ref{tab:bkpipicontrib} 
with the respective tree and penguin contributions, followed by CKM weak phases
contained in the amplitude. 
Since magnitudes and phases
extracted by the fit procedure are the overall contributions, it's not possible to isolate and measure 
$\gamma$ in a simple amplitude analysis of $B^+$ and $B^-$ and $B^0$ and $\bar{B^0}$.

{\small
\begin{table}[!h]
  \begin{center}
\begin{tabular}{c|c|c|c}
\centering
 & resonance & contribution & weak phase\\
\hline \hline
\multirow{5}{*}{$B^+ \to$} & $K^*(890)^0 \pi^+$ & \multirow{2}{*}{$V_{bt}V^*_{ts} \; P$} &
\multirow{2}{*}{} \\ \cline{2-2}
& $K^*(1430)^0 \pi^+$   & \\ \cline{2-4}
& $K^+ \rho(770)^0$ & \multirow{2}{*}{$V_{bt}V^*_{ts} \; P + V_{bu}V^*_{us} \; T^C_S$} &
\multirow{2}{*}{$\gamma$} \\ \cline{2-2}
& $K^+ f_0(980)$ & & \\ \cline{2-4}
& $K^+ \chi_{c0}$ & $V_{bc}V^*_{cs} \; T_S$ &   \\ \hline \hline
\multirow{5}{*}{$B^0 \to$} & $K^*(890)^+ \pi^-$ & \multirow{2}{*}{$V_{bt}V^*_{ts} \; P + V_{bu}V^*_{us} \; T^C$} &
  \multirow{2}{*}{$\gamma$} \\ \cline{2-2}
& $K^*(1430)^+ \pi^-$ &  &  \\ \cline{2-4}
& $K_S \, \rho(770)^0$ & \multirow{2}{*}{$V_{bt}V^*_{ts} \; P + V_{bu}V^*_{us} \; T_S$} &
 \multirow{2}{*}{$\gamma$} \\ \cline{2-2}
& $K_S \, f_0(980)$ & & \\ \cline{2-4}
& $K_S \, \chi_{c0}$ & $V_{bc}V^*_{cs} \; T_S$  \\
\hline \hline
\end{tabular}
\end{center}
	\caption{Resonances and dominant contributions from $B \to K \pi \pi$ decays . We denote
    allowed, suppressed by color and penguin amplitudes by $T^C$, $T_S$ and $P$.}
    \label{tab:bkpipicontrib}
\end{table}
}

Based in SU(3) flavor symmetry, we expect the same \emph{penguin} amplitudes
for the four processes $B^\pm \to K^* \pi^\pm$ and $B^0,\bar{B^0} \to K^{*\pm} \pi^\mp$.
The method consist to extract the $B^\pm \to K^* \pi^\pm$ penguin parameters
\footnote{For simplicity we refer only to $K^*$, while in practice the method uses simultaneously the parameters of the
resonances $K^*(890)$ and $K^*(1430)$.}, introduce them in the $B^0$ and $\bar{B^0}$ amplitudes,
then use a new technique of joint fit to extract the tree phases, where $\gamma$ can be obtained from.

As the magnitudes and phases are measured relative to a fixed resonance, the compatibility
between penguin parameters from $B^\pm$ and $B^0,\bar{B^0}$ to $K^* \pi$, requires
that the amplitude analysis be made relative to an anchor resonance that have the same amplitude for $B$ charged and
neutral. We use the non CP violating $B \to K \, \chi_{c0}$ amplitude.
The asymmetry measured by Belle \cite{bellechi} for the channel $B^\pm \to K^\pm \chi_{c1}$ is 
$A_{CP}=-0.01 \pm 0.03 \pm 0.02$, indicating as expected, 
that the dominant contribution is a tree diagram without weak phase.
The equality between $B$ charged and neutral amplitudes, is based in flavor SU(3) symmetry considerations.

Following, there is a schematic representation of the method, where arrows represent quantities extracted by fit
and $\theta^\pm=(\delta_T \pm \gamma)$:
\begin{flushleft}
\begin{eqnarray}
B^\pm \to K^{*0} \pi^\pm \quad  : \quad a_P \, e^{i \, \delta_P} \hspace{1,7cm}	\nonumber \\
\Downarrow \hspace{2,1cm} \nonumber \\
B^0 \to K^{*+} \pi^-   \quad  \,  : \quad  a_P \, e^{i \, \delta_P} + a_T \, e^{i \, \theta^+} \nonumber \\
\bar{B^0} \to K^{*-} \pi^+  \quad : \quad  a_P \, e^{i \, \delta_P} + a_T \, e^{i \, \theta^-} \nonumber \\
\Downarrow \hspace{0,5cm} \nonumber \\
\gamma = \frac{\theta^+ - \theta^-}{2}.
\label{eq:esquemag}
\end{eqnarray}
\end{flushleft}

The method is based in three basic and well accepted hypothesis that can be tested during the analysis. First: the dominant
contribution of $B^\pm \to K^* \pi^\pm$ is $V_{bt}V^*_{ts} \; P$. This has been partially confirmed by BaBar \cite{BaBark2pi}.
The test, is to check if the $B^+ \to K^* \pi^+$ and $B^- \to K^* \pi^-$ amplitudes are the same.
Second: the penguin components from $B^\pm \to K^* \pi^\pm$ and $B^0, \bar{B^0} \to K^{*\pm} \pi^\mp$
are equal \cite{teo-chiang, teo-beneke}.
Third: $\chi_{c0}$ have the same amplitude for $B^\pm \to K^\pm \chi_{c0}$ and
$B^0, \bar{B^0} \to K_S \, \chi_{c0}$. 
The experimental test for the second and third hypothesis, consists in the equality of the tree magnitudes extracted
from the intermediary process $B^0$ and $\bar{B^0}$ to $K^{*\pm} \pi^{\mp}$.
The confidence in the result of $\gamma$ is based in the fulfillment of all hypothesis,
whereas the failure of one implies in interesting unexpected effects.

\section{Amplitude Analysis}
To extract the $B^\pm \to K^\pm \pi^+ \pi^-$  parameters, we apply a maximum likelihood fit to the isobaric amplitudes
\begin{equation}
\mathcal{A}^\pm= a_\chi e^{i \, \delta_\chi}  \, \mathcal{A}_\chi + \sum_i \, a_i^\pm e^{i \, \delta_i^\pm} \, \mathcal{A}_i,
\label{eq:ampanalise}
\end{equation}
where $\mathcal{A}_i$ are functions of Dalitz variables modeled by Breit-Wigner distributions times angular functions and form
factors; $i=K^*(890)^0,\, K^*(1430)^0,\, \rho(770)^0,\, f_0(980)$
and $a_\chi, \delta_\chi$ parameters are fixed.

Regarding the neutral system, the separation between $B^0$ and $\bar{B^0}$ is not obvious, having in sight the final state
$K_S \pi^+ \pi^-$. In addition, there is mixing among $B^0$ and $\bar{B^0}$, introducing time dependence in the
probabilities. One possibility is to use tagging for creating two separated sets of events and apply a maximum likelihood
fit using 
\begin{eqnarray}
\! M(\Delta t)=e^{-(\Gamma /2 - i M) \Delta t} \, [ \mathcal{A} \, \cos{(\Delta m \, \Delta t/2)}  \nonumber \\
\qquad -i \, q/p \, \bar{\mathcal{A}} \, \sin{(\Delta m\, \Delta t/2)} ], \\ 
\bar{M}(\Delta t)= e^{-(\Gamma /2 - i M) \Delta t} \, [\bar{\mathcal{A}} \, \cos{(\Delta m \, \Delta t/2)} \nonumber \\
-i \, p/q \, \mathcal{A} \, \sin{(\Delta m\, \Delta t/2)}],
\end{eqnarray}
where $\mathcal{A}$ e $\bar{\mathcal{A}}$ are time-independent amplitudes analogue to (\ref{eq:ampanalise}),
for $B^0 \to K_S \pi^+ \pi^-$ and $\bar{B^0} \to K_S \pi^+ \pi^-$ decays.

The probability distribution for the final state $K_S \pi^+ \pi^-$, independent of it's origin, is given by the sum
$|M(\Delta t)|^2+|\bar{M}(\Delta t)|^2$. It was found \cite{burdman, gardner} that in the case $|p/q|=1$, that this sum displays
the interesting property of canceling mixing and time dependence terms,
\begin{equation}
|M(\Delta t)|^2+|\bar{M}(\Delta t)|^2 = e^{-\Gamma \, t} \, ( \, |\mathcal{A}|^2+|\bar{\mathcal{A}}|^2 \, ).
\label{eq:somaamps}
\end{equation}
For the neutral system, we can use the total normalized amplitude
\begin{eqnarray}
\frac {  | a_\chi e^{i \, \delta_\chi} \, \mathcal{A}_\chi + \sum_i \, a_i e^{i \, \delta_i} \, \mathcal{A}_i|^2  
+ | \, \bar{a_\chi} e^{i \, \bar{\delta_\chi} } \, \mathcal{A}_\chi + \sum_i \, \bar{a_i} e^{i \, \bar{\delta_i} } 
\, \mathcal{A}_i|^2  } {N_1 + N_2} , \nonumber
\label{} 
\end{eqnarray}
\begin{eqnarray}
N_1 = \int |a_\chi e^{i \, \delta_\chi} \, \mathcal{A}_\chi + \sum_i \, a_i e^{i \, \delta_i} \, \mathcal{A}_i|^2 \, ds_{ij} \, ds_{jk}, \nonumber \\
N_2 = \int | \bar{a_\chi} e^{i \, \bar{\delta_\chi} } \, \mathcal{A}_\chi + \sum_i \, \bar{a_i} e^{i \, \bar{\delta_i} }
\, \mathcal{A}_i|^2 \, ds_{ij} \, ds_{jk}.
\label{eq:amptotmista}
\end{eqnarray}

As a fundamental step in $\gamma$ extraction, we developed a novel method of \emph{joint fit} for extracting the parameters
$a_i, \delta_i$ and $\bar{a_i}, \bar{\delta_i}$ from the same set of untagged time-independent events \cite{bgm}.
In this method, we apply a maximum likelihood fit using as PDF the eq.(\ref{eq:amptotmista}),
were three parameters $a_\chi, \delta_\chi$ and $\bar{\delta_\chi}$ are fixed.
Despite the fact that $\chi_{c0}$ resonance parameters are equal, the term $\bar{a_\chi}$ is kept free instead of
being fixed to the same value of $a_\chi$. This should be interpreted as if we multiplied the amplitude 
$\bar{\mathcal{A}}$ by a \emph{scaling} parameter $s$. In this case, $\bar{a_\chi}= s \, a_\chi$ whereas the other
parameters $\bar{a_i}$ contains $s$ also. That allows one to measure difference in number of events from 
$B^0$ and $\bar{B^0}$, investigating direct CP asymmetry.
In one single procedure it's possible to extract in an
independent way, the parameters from the amplitudes $\mathcal{A}$ and $\bar{\mathcal{A}}$
and the ratio in number of events, calculated by the ratio of the normalizations $N_1/N_2$.

\begin{figure}[hbt]
\centering
\includegraphics[scale=.35]{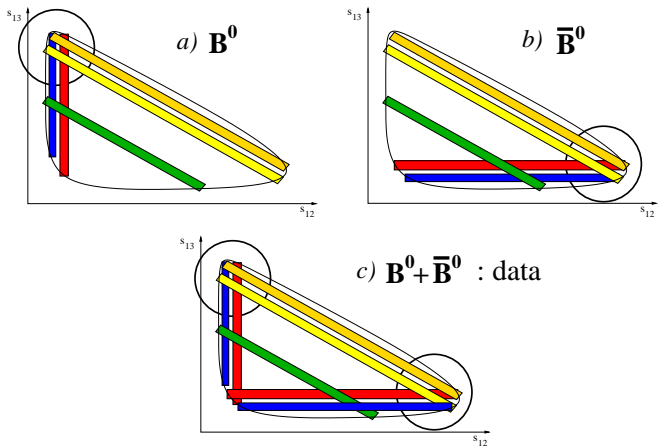}
\caption{Schematic representation for the stand-alone DP of $a) B^0 \to K_S\pi^+\pi^-$, $b) \bar{B^0} \to K_S\pi^+\pi^-$ and in
$c)$ the joint DP obtained from the untagged analysis. Bands are pictorially representing resonances regions, where for simplicity
we ignore angular effects in the distribution. Circles are displaying regions of interference.
Is the non-overlapping of circles that assures the separation between both amplitudes in the joint fit analysis.}
\label{fig:ampsdados}
\end{figure}

In the convention $(K_S, \pi^+, \pi^-)\to (1,2,3)$ for the final state particle numbering,
the charge conjugation operation switches the Dalitz variables $s_{12} \leftrightarrow s_{13}$.
The resonances $K^*$ bands are centered in different axis for $\mathcal{A}$ and $\bar{\mathcal{A}}$,
establishing some sort of signature for the event origin $B^0$ or $\bar{B^0}$, as can be seen in figure 
\ref{fig:ampsdados}. However, in the untagged analysis, data supplies a joint Dalitz plot, where a generic point
have an unknown origin. Although we cannot distinguish events, the joint fit can 
identify two different overlapping surfaces of $B^0$ and $\bar{B^0}$ due to the non-overlapping of at least one
interference region from both amplitudes. The non-overlapping
assures that the amplitudes can be explored in an independent way by the fitting procedure,
guaranteeing the unicity in the fit result.

In the Belle \BdtoKpipi analysis 
\cite{bellekspipi}, the authors fit a untagged sample of $K_S \pi \pi$, using as total
amplitude the sum $|A(x,y)|^2 + |A(y,x)|^2$. The amplitudes have same parameters of magnitude and phase,
assuming no CP violation and differing only by a Dalitz variable exchange.
Using the joint fit technique, it's possible to extract different parameters to both decays and
measure experimentally CP violation, giving one step ahead in the technical difficulty existent so far.


\section{Feasibility Study}
\begin{figure}[hbt]
\centering
\includegraphics{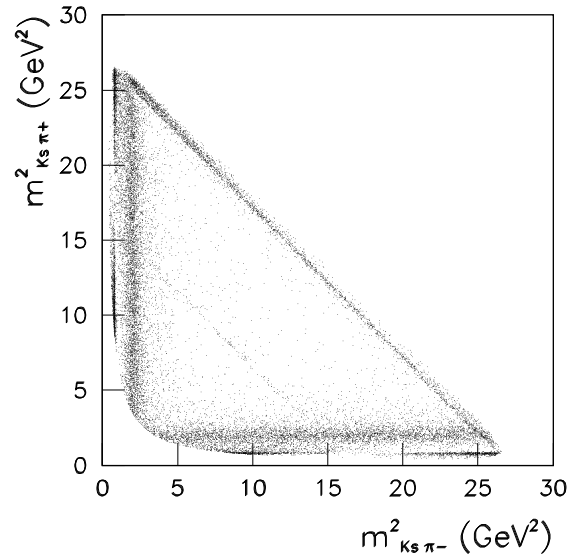}
\caption{$B^0$, $\bar B^0 \to K_s \pi^+ \pi^-$ events distribution from fast MC generated with table 
\ref{tab:viabilidade} parameters.}
\label{fig:dalitzviab}
\end{figure}

To investigate the feasibility and the error dimension in the joint analysis, we generated and fitted
100 Monte Carlo experiments of 100K \BdtoKpipi events (the expected number 
for one year of data taking in LHCb \cite{lhcb}).
All resonances are described by Breit-Wigner distributions and the total amplitude is given by (\ref{eq:amptotmista}).
The parameters used in the generation were inspired in the observed parameters by \cite{BaBark2pi, bellek2pi}.
The Dalitz plot distribution for one generated experiment of 100K events is displayed in figure 
\ref{fig:dalitzviab}, where the small contribution of the resonance $\chi_{c0}$ can be seen.

In table \ref{tab:viabilidade}, column input, we show the generated magnitudes and phases 
to each resonance of $B^0$ and $\bar{B^0}$. 
The extracted parameters by the joint fit are shown in the last column, where the displayed values are
the central values of the gaussian distributions of the 100 experiments and the errors are the
width of the same gaussian distribution. One can note that the extracted quantities are in agreement
with the generated ones and with small errors, stating the feasibility of the joint fit method.
\begin{table}[htb]
\begin{center}
  \begin{tabular}{|c c|c|c|}
    \hline
decay &$B^0/\bar B^0$ & input &  fit 100K events  \\
    \hline
$\chi_c K_s$ & $a_0/\bar a_0$ & 0.30/0.30 &  fixed/$(0.30 \pm 0.03)$  \\
& $\delta_0/\bar\delta_0 $ & 3.78/3.78  & fixed/fixed \\
 \hline
$K^*(890)\pi$ & $a_1/\bar a_1$ & 1.17/1.30 & $1.17 \pm 0.06 $  /$(1.30 \pm 0.01)$ \\
 & $\delta_1/\bar \delta_1$ & 0.40/5.98 &  $(0.41 \pm 0.08)$/$(5.99 \pm 0.07)$  \\
 \hline
$K_0^{*}(1430)\pi$ & $a_2/\bar a_2$ & 2.45/2.72 &  $(2.45 \pm 0.11)$/$(2.72 \pm 0.13)$  \\
 & $\delta_2/\bar \delta_2$ & 0.375/6.00 &   $(0.39 \pm 0.08)$/$(6.00 \pm 0.06)$ \\
 \hline
$\rho^0 K_s$ & $a_3/\bar a_3$ & 0.60/0.60 &   $(0.60 \pm 0.04)$/$(0.60 \pm 0.04)$ \\
& $\delta_3/\bar \delta_3$ & 1.20/1.20  &  $(1.22 \pm 0.09)$/$(1.20 \pm 0.07)$ \\
 \hline
$f_0 K_s$ & $a_4/\bar a_4 $ & 1.03/1.03 &  $(1.02 \pm 0.06)$/$(1.04 \pm 0.05)$  \\
& $\delta_4/\bar \delta_4$ & 2.30/2.30 &   $(2.30 \pm 0.07)$/$(2.30 \pm 0.08)$ \\
 \hline
 \end{tabular}
  \begin{tabular}{|c |}
  $N^0( B^0 \to K_s \pi^+ \pi^-)/ N^0(\bar B^0 \to K_s \pi^+ \pi^-)$  =
  0.84 $\pm$  0.12  \\
 \hline
  \end{tabular}
  \caption[]{ Monte Carlo simulation for $B^0$,   $\bar B^0 \to K_s \pi^+ \pi^-$ 
  decay. We generated sample with the the parameters $a_i$ and $\delta_i$ for 
  $B^0$ and $\bar a_i$ and $\bar \delta_i$ for $\bar B^0 $. The third 
  column display the fit results with one hundred samples for each  
  100K events. The last line is the ratio between $B^0, \bar B^0 \to K_s \pi^+ \pi^-$
  number of events.}
  \label{tab:viabilidade}
\end{center}
\end{table}

One important issue concerning the $\gamma$ extraction, is the size of the ratio $r=a_T/a_P$ and the phase
difference $\theta=\delta_T - \delta_P$ from the tree and penguin amplitudes of the $K^*$ resonance.
The theoretical knowledge of these quantities is model dependent.
Some groups using factorization approach \cite{neubert}, arrive to large $r$ and small $\theta$.
On the other hand, non-factorisable approach for pseudoscalar-pseudoscalar $B$ decay \cite{buras},
presents an opposite scenario, with small $r$ and large $\theta$. The joint fit applied to the
real data, will be able to define which theoretical approach is more adequate, 
since it is possible to measure the parameters under discussion. 
To our study, we choose $r=0.45$ and generate the experiments with $\gamma=69^\circ$.

If in the unconstrained analysis, the three hypothesis prove to be correct, we can assume that the $K^*$
resonances parameters from $B^0$ and $\bar{B^0}$ are given by the following tree and penguin sum:
\begin{eqnarray}
a \, e^{i \, \delta}= a_P \, e^{i \, \delta_P} + a_T \, e^{i \, \theta^+}, \nonumber \\
\bar{a} \, e^{i \, \bar{\delta}}= a_P \, e^{i \, \delta_P} + a_T \, e^{i \, \theta^-}.
\label{eq:pingtreekstar}
\end{eqnarray}
In this case, we can use the scheme (\ref{eq:esquemag}) to extract $\gamma$ using a constrained
fit, where the penguin amplitude is fixed and the tree parameters $a_T$ and $\theta^\pm$ 
are directly extracted from $K^*$.
Using the parameters from table \ref{tab:viabilidade}, we measure $\gamma= 69^\circ \pm 5^\circ$.


\section{Conclusion}
 
We presented a method to extract the CKM $\gamma$  angle using a 
combined Dalitz plot analysis from $B^{\pm} \to K^{\pm} \pi^+ \pi^-$ 
and  $B^0, \bar B^0 \to K_s \pi^+ \pi^-$. 
This approach use three basic hypothesis that can be tested before one proceeds to the constrained fit.
For measuring the $B$ neutral parameters, we use a new technique of joint fit
that allow us to extract in an independent way the amplitudes from two summed surfaces in 
a joint sample of $B^0 + \bar B^0$ untagged events. 
We carried a fast MC study to estimate the errors associated to the joint fit technique. 
Assuming a relatively low
statistics for the anchor resonance $\chi_{c0}$ we obtained $\gamma$ with a $5^\circ$ error.

During the analysis we can measure CP violation by counting the number of events from $B^0$ and $\bar{B^0}$,
information extracted in the joint fit,
or exploring Dalitz symmetries as discussed in \cite{burdman}.
We can also measure the ratio and phase difference from tree and penguin amplitudes of the $K^*$ resonance, 
defining which theoretical approach, factorisable or non-factorisable is more adequate.

The method presented here is competitive with the 
other approaches to determine the CKM  $\gamma$ angle \cite{ schneider}
and needs   the high  statistics expected for the LHCb experiment, 
due to the  small contribution of the reference channel $B\to \chi^0 K$. 
However, in the case  that the  $ f_0(980) $ resonance is dominated by the $ s \bar s$ component
\cite{bnn},   or even if the ratio between the tree and penguin is  
negligible \cite{f0},  the $B\to f_0(980) K$ amplitude could take  place 
of the charmonium as a reference channel in the analysis.

\end{document}